# MECHA:다중스레드 및 효율적인 암호화 하드웨어 액세스

프라타마 데리 [1], 락스모노 아구스 마하르디카 아리 [2], 이크발 무함마드 [2], 김호원 [3]

[1] 부산대학교 정보융합공학과 박사과정
[2] 부산대학교 정보융합공학과 석사과정
[3] 부산대학교 정보융합공학과 교수

derryprata@gmail.com, agusmahardika@pusan.ac.kr, iqbal@islab.re.kr, howonkim@pusan.ac.kr

# MECHA: Multithreaded and Efficient Cryptographic Hardware Access

Pratama Derry, Laksmono Agus Mahardika Ari, Iqbal Muhammad, Howon Kim
Dept. of Computer Information Convergence Engineering, Pusan National University

## Abstract

This paper presents a multithread and efficient cryptographic hardware access (MECHA) for efficient and fast cryptographic operations that eliminates the need for context switching. Utilizing a UNIX domain socket, MECHA manages multiple requests from multiple applications simultaneously, resulting in faster processing and improved efficiency. We comprise several key components, including the Server thread, Client thread, Transceiver thread, and a pair of Sender and Receiver queues. MECHA design is portable and can be used with any communication protocol, with experimental results demonstrating a 83% increase in the speed of concurrent cryptographic requests compared to conventional interface design. MECHA architecture has significant potential in the field of secure communication applications ranging from cloud computing to the IoT, offering a faster and more efficient solution for managing multiple cryptographic operation requests concurrently.

## 1. Introduction

Ensuring secure communication has become an increasingly important issue in our connected world. The speed and efficiency of cryptographic operations are crucial to the success of these endeavors. In this paper, we present a multithread and efficient cryptographic hardware access (MECHA), an application programming interface (API) architecture that significantly improves the performance of cryptographic operations by managing multiple requests from multiple applications simultaneously, all while eliminating the need to switch context between each cryptographic operation request.

Conventionally, APIs use context switching to share the usage of a single cryptographic hardware, resulting in slower processing times. Each of the applications that shares the same cryptographic hardware needs to wait for other application requests to be finished. Once finished, the API will switch the context and acquire the crypto hardware handle.

We utilize an UNIX domain socket, which enables us to manage multiple cryptographic operation requests from various applications at once, without the need for context switching. This results in faster and more efficient processing of cryptographic operations.

There are several Parallel Programming Languages and API models that can be implemented [1]. Our MECHA consists of several key components, including the Server thread, Client thread, Transceiver thread, Sender and Receiver queues. The Server thread is created by the application requesting cryptographic operations, while the Client thread is created by the Server thread for each application that requires cryptographic operations. Multiple Client threads are saved in a thread pool, which puts the requested operation and response in a Sender queue and Receiver queue. The Transceiver thread is inside the Server thread and schedules the sending of data from the Send queue to the crypto hardware and puts it into the Receiver queue.

Since the interfacing layer is designed portable, MECHA can be used with any communication protocol, including SPI, UART, I2C, and others. Experimental results demonstrate a remarkable improvement in performance, with 82.8% times increase in the speed of concurrent cryptographic requests compared to conventional interface design. MECHA offering a faster and more efficient solution for managing multiple cryptographic operations simultaneously.

## 2. Theoretical Background

### 2.1 Hardware Security Module

A Hardware Security Module (HSM) is a specialized device designed to provide secure key storage, cryptographic processing, and key management functions. HSM, also known as cryptographic accelerators, allows for fast cryptographic operations and secure cryptographic key management [2]. They employ a range of physical and logical security mechanisms to prevent unauthorized access, tampering, and extraction of keys and other sensitive data. HSMs are widely used in industries such as banking, government, and healthcare, where data security is of paramount importance. In this paper, the HSM refers to Crypto Hardware.

### 2.2 Multithreading

Multithreading is a programming technique that allows multiple threads of execution to run concurrently within a single program. Each thread operates independently, allowing for efficient use of system resources and faster program execution. It is commonly used in applications that require a high degree of parallelism. However, multithreading also introduces new challenges, such as race conditions, deadlocks [3], thread safety and security [4] issues. To fully exploit the benefits of multithreading, a thorough understanding of the underlying principles and best practices is essential.

### 2.2 Unix Domain Socket

A Unix domain socket (UDS) is a communication endpoint that allows processes to exchange data on the same host [5]. Unlike network sockets, UDS operates entirely within the operating system kernel, providing high performance and low latency communication between processes. UDS are widely used in Unix-like operating systems for inter-process communication (IPC), and they are a popular choice for local client-server architectures. Their advantages over other IPC mechanisms are pipes and message queues, including reliability, scalability, and flexibility.

### 2.3 Cryptographic Library

A crypto library is a software library that provides cryptographic functions, such as encryption, decryption, hashing, and digital signature generation and verification [6]. OpenSSL, mbedTLS, gnuTLS, and wolfSSL are used to implement secure communication protocols, authenticate users, and protect data confidentiality and integrity. They provide a set of standardized algorithms and protocols, such as AES, RSA, and TLS, and offer a high-level interface for application developers to use. Crypto libraries can also provide additional features, such as key management, random number generation, and password hashing, to support the development of secure applications.

## 3. Proposed Design

### 3.1 Design Overview

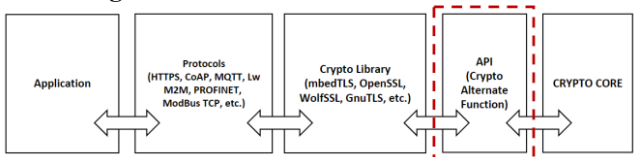

Figure 1. Crypto Hardware API Stack

In this paper we implement MECHA for alternate crypto functions by using broker server. In conventional design, the communication between application with crypto core is done through crypto library as seen as in Figure 1. In OpenSSL this part is called crypto engine, while in mbedTLS it is called as alternate crypto function.

Traditionally, this kind of communication will use a context switch mechanism through the API for different use of cryptographic function calls. For example, when an application #1 ($A^1$) is calling hash function, then another application #2 ($A^2$) is calling block cipher encryption function, the API will send the hash instruction first and fetch the response, while this happens, the $A^2$ will wait until the context is free. Finally, after the $A^1$ request is done, the context is free, and API will forward block cipher instruction from $A^2$ and get the response from crypto core. This sequence will cause bottleneck in the API context queue.

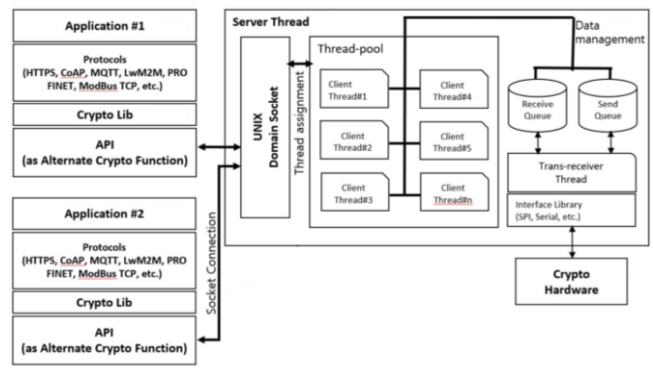

Figure 2. MECHA Architecture Overview

Therefore, we proposed a multithread and efficient cryptographic hardware access architecture (MECHA). Where each of cryptographic request from $A^n$ will be received by the Server Thread (ST) created by each of the applications that uses the crypto hardware module. Only one ST exist in an operating system, when an application tries to connect to ST and success, it does not need to create an ST.

### 3.2 Server Thread

The creation of ST depends on the priority configuration, we provide a configuration with lists of application with corresponding priority to use the crypto hardware. The application that has the most priority will be the one that creates a server thread. If there is no configured priority, by default our system will prioritize the first request, first-come-first-served basis.

The ST consists of a UDS to receive cryptographic requests from other applications. For each source, a Client Thread (CT) is created in a thread-pool see Figure 2. When the request is received, the ST will classify the request from different applications based on Connection Socket Number (CSN) by creating CT for each application. The classification process in CT is done by appending the CSN as prefix of the Protocol Data Unit (PDU) packet in Figure 4. Once the packet is appended, server will put the packet into a Send Queue (SQ). SQ will buffer all the requests from $CT^0$ to $CT^n$. If Transceiver Thread (TT) finished with the last request / ready signal from crypto core, it will send all the PDU packets saved in SQ. This is where our design improves the transmission efficiency, TT will forward all the packets from different applications with the currently opened crypto core context without needing to wait for the context to free from previous usage.

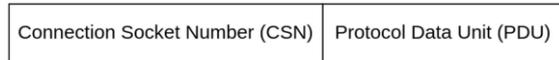

Figure 3. CSN is appended as PDU prefix

Assuming that the crypto core is ready to retrieve appended PDU packets, the response is a form of FIFO packets which are saved in receiver queue (RQ). Each CT will then pick each of it owns response from crypto core by CSN matching. If a response is found, CT will pass the response to corresponding application socket via UDS.

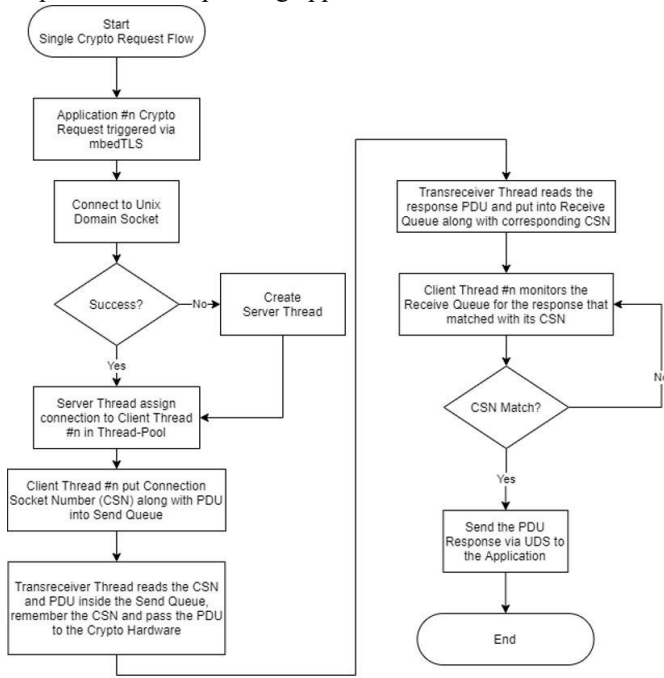

Figure 4. A single cryptographic operation request flow

Using this approach, multiple applications ($A^n$) will be able to access and request cryptographic operation concurrently without sacrificing bottleneck in context switching caused by connection re-initiation with crypto hardware.

## 4. Result and Analysis

We implement MECHA prototype with an FTDI with MPSEE SPI mode connection between the application that is running on Ubuntu Linux and the crypto hardware is FPGA that has loopback SPI access. Our experiment benchmarks the performance of multiple $n$ applications requesting different cryptographic operation concurrently in one hardware. A set of PDU is sent with fixed length in a loop and when all responses are received, the request is assumed to be done.

In Figure 5, we benchmark the performance of MECHA receiving multiple cryptographic requests concurrently, each application instance is transmitting data with the length of 64KB at the same time. At first with 5 instances running concurrently, only 1.91 seconds faster than conventional context switching API. However, as the instance access increased, the time that context switch API took to complete all the request also increased linearly. At 80 instances, which meant the transferred data is around 5.12MB, MECHA performs 82.8% faster than the conventional context switch API, thanks to the PDU management in server thread that combines the data before batch sending instead of waiting for each application access context is finished.

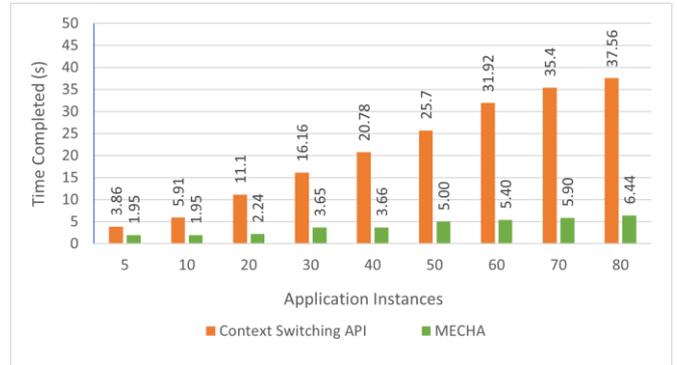

Figure 5. MECHA performance benchmark

## 5. Conclusion

We propose MECHA for cryptographic hardware interface, by reusing the context to optimize the transmission efficiency and multithreading architecture for data management, achieving 82% faster data transmission for cryptographic operation request than conventional interface in our experiment. This proves that MECHA is feasible and can replace the conventional context switching API design for crypto hardware interfaces, offering a faster and more efficient solution for managing multiple cryptographic operations simultaneously in one cryptographic hardware.

## 6. Acknowledgement

This work is financially supported by Korea Ministry of Land, Infrastructure and Transport (MOLIT) as 「Innovative Talent Education Program for Smart City」